\documentclass[conference]{IEEEtran}
\usepackage{stmaryrd}
\usepackage{amsfonts}
\usepackage{graphicx,times,amsmath} 

\IEEEoverridecommandlockouts    

\begin{document}

\title{\ \\ \LARGE\bf On the use of Stress information in Speech for Speaker Recognition
\thanks{Laxmi Narayana M and Sunil Kumar Kopparapu are with TCS Innovation Lab
- Mumbai, Tata Consultancy Services, Yantra Park, Thane (West) Mumbai, Maharastra, India.  
(email: m.laxminarayana@gmail.com, sunilkumar.kopparapu@tcs.com)}}

\author{Laxmi Narayana M and Sunil Kumar Kopparapu}


\maketitle

\begin{abstract}
The performance of a speaker recognition system decreases when the speaker is under 
stress or emotion. In this paper we explore and identify 
a mechanism that enables use of 
inherent stress-in-speech or speaking style information present in speech of a
person as additional cues for speaker recognition. 
We quantify the 
the inherent stress present in the speech of a
speaker mainly using $3$ features, namely, 
pitch, amplitude and duration (together called PAD) 
We experimentally observe that the PAD vectors of similar phones in different
words of a speaker are close to each other in the three dimensional (PAD) space 
confirming that the way a speaker stresses different syllables in their 
speech is unique to them, thus we  
propose the use of PAD based speaking style 
of a speaker as an additional feature for speaker recognition applications. 


\end{abstract}


\section{Introduction}
\label{sec:introduction} \vspace{-4pt}

\PARstart{S}{everal} speech features for speaker recognition 
and speech recognition applications have been proposed in literature.
Mel Frequency Cepstral Coefficients (MFCC) \cite{merm80} by far, have been 
the most commonly used 
speech features \cite{reyn95} \cite{reyn295} \cite{rash04} \cite{sedd04} \cite{reyn00}. 
Other speech features like Linear Frequency Cepstral coefficients (LFCC)
\cite{merm80}, 
wavelet octave coefficients of residues (WOCOR) \cite{neng07} are also 
used for speech and speaker recognition applications and good performance of those systems have also been reported. 
Combination of two or more speech features for speaker recognition applications is also in practice and the literature reports an 
improved performance with combination of multiple speech features \cite{neng07}. 
Nevertheless automatic speech and speaker recognition systems function less
efficiently when the speaker is under stress or emotional state.   
In this paper, we assume that the way a person expresses his emotion and 
stresses certain syllables in their speech 
is unique to them and this style of speaking is consistent for a speaker. 
We make this assumption based on observations and this motivates 
the current work.  
We further explore to identify a mechanism to use the inherent stress 
information in the speech 
as an additional feature for speaker or speech 
recognition\footnote{These features are in addition to the traditionally used speech
features}.

Prosody is the melody of speech \cite{quat04} and 
emotion-initiated gestures and stressed syllables in human speech communication
help 
in the improvement of speech understanding. 
However, stress and emotional expressions in human speech had been found to be the source of difficulty 
for some applications like automatic speech recognition (ASR) and 
speaker recognition (SR). 
The reason behind the manifestation of stress in human speech is due to several factors. 
The first and foremost reason is that emphasis of some syllables in human speech is natural and this becomes 
very prominent when the speaker is emotional or under stress. 
Many studies that consider stress in speech as distortion introduced by emotion
have shown that these factors can 
severely reduce speech recognition accuracy. 
Techniques for detecting or assessing the presence of stress could 
help neutralize stressed speech and improve 
robustness of speech recognition systems. 
Although some acoustic variables derived from linear speech production theory have been investigated as 
indicators of stress, they are not consistent.

Milan \cite{milan09} mentions in his work on spectral analysis of stressed speech that stress is a 
psycho-physiological state characterized by subjective strain, 
dysfunctional physiological activity and deterioration of performance. 
The accepted term for speech signal carrying information on the speaker's 
physiological stress is `stressed speech'. 
This refers to the imprints of stress in the speech of person when under  
stress. 
In other words, the stress experienced by the speaker is reflected in their speech. 
In contrast to this \cite{stress_labeling} reports the work of stress 
labeling of syllables, 
analyzes the stress in normative speech - uttered in a neutral speaking style or when the 
speaker is not stressed. 
In this case, the `stressed speech' refers to that portion of speech, 
which is stressed naturally with no influence of the speaker's mental or 
psycho-physiological state. 

As mentioned earlier, our motivation for using the natural `stress' information in human speech as a feature for speaker recognition is based on the 
assumption that the {\em speaking style} of a speaker is consistent and unique to the speaker. 
One can 
also extend this idea to say that the people of same geographical area have a similar speaking style (or accent). 
{\em If the way a speaker speaks is unique to the speaker, then why not use
that `speaking style' information specifically 
as an additional feature for speaker recognition?} 
We interpret that the speaking style of a speaker is a reflection of the way 
the speaker stress certain syllables in a sentence or phrase or a word. 
We translate that the speaking style of a speaker can be parametrized by 
identifying the parameters of {\em stress} in speech.

The rest of the paper is organized as follows. 
Section \ref{sec:theory} gives a summary of literature on the stress related 
features used in speech. 
Section \ref{sec:approach} discusses our approach of using the inherent stress information in speech for speaker recognition. 
Section \ref{sec:experiments} gives the details of the experiments conducted to quantify the speaking style of a speaker. 
and we conclude in Section \ref{sec:conclusions}.


\section{Stress related features in Speech}
\label{sec:theory}

Stress and its manifestation in the acoustic signal have been the subject 
matter of many studies 
in literature \cite{stress_labeling} \cite{jupiter} \cite{unstressed}. 
Researchers have attempted to determine reliable indicators of stress 
by analyzing certain variable parameters of speech such as fundamental 
frequency (pitch), 
amplitude, concentration of spectral energy, duration and several 
others \cite{unstressed} \cite{wight} \cite{min} \cite{telugu} \cite{cues}. 
In literature, analysis of stress is done through  
analysis of some parameters of stress like fundamental frequency (F0), 
pitch, vowel duration and formants in recorded emotional speech, namely,
analyzing a speaker's speech 
when they are under stress, fatigue, heavy workload, environmental noise, 
sleep loss or expressing some emotion like happiness, anger or sorrow. 

Literature clearly distinguishes the speech as 
(a) uttered when the speaker is under stress or expressing some emotion and 
(b) in a neutral speaking style, namely, when the psychological state of the 
speaker doesn't seriously affect the speech, 
for example, reading news or even normal conversations. 
To our best knowledge the available literature talks  
only about the speech of type (a) to analyze stress in speech.
The inherent presence of stress in some syllables of speech whether it is emotional or normative, 
is natural in accordance with the influencing factors like language, accent and the geographical location to which a speaker belongs to. 
If the stress is absent in non-emotional speech, the corresponding pitch and
amplitude contours would have been absolutely flat. 
We believe that the manifested stress in speech whether the speaker is under stress or in normal condition, 
is distinguished by the intensity of the parameters of speech (or stress), but, 
nevertheless, the set of parameters that quantify stress is same in both the cases. 
A speaker naturally and unintentionally stresses some syllables while speaking; 
there exists an inherent mechanism behind this unintentional occurrence of stress in speech, 
which is unique to a person or people belonging to a geographical region. 
We believe that this information can be used as an additional feature in speaker recognition applications. 
We intend, in this paper, to  study the parameters of stress and seek 
to use the 
stress information in normative speech in speaker identification/verification. 
We summarize  our study of literature on how 
`stress' in speech is parameterized. 

Higher intensity, greater duration and higher F0 are believed to be the primary acoustic cues 
for stressed syllables, although how the three factors work together to make a syllable more prominent than the surrounding ones is still not very clear. 
Therefore, these cues are used as the main acoustic features in the stress 
detection task in some studies (example, \cite{wight}). 
Stressed syllables are usually indicated by high sonorant energy, long syllable or vowel duration
and high and rising F0 is cited by \cite{jupiter}. 
Stress is found to be correlated with voice quality as well. 
Usually, stressed vowels are pronounced clearer and unstressed vowels tend to 
have reduced clarity. 
When listening to an utterance, people not only use acoustic cues but also 
syntactic and/or semantic cues to help the location of stress. 
Therefore, features derived from the text, such as part of speech (POS) and the position in the phrase are as well used for detection of 
stress \cite{min}. 
Many works in literature mention that the  assignment of stress is based on a 
relative comparison of the syllables within a word and 
does not rely on a global model of a stressed or unstressed syllable.

Stress, accent and/or emphasis detection all deal with the detection of the relative prominence of a syllable within a word. A discussion of the different detection methods is difficult, because the word "stress" has been used ambiguously to refer to several types of prominence, including strong vs. weak syllable distinctions as indicated by 
lexical stress marking, as well as phrasal prominence as indicated by a pitch accent \cite{min}.
The cues for stress discussed above have been found mainly in English and Dutch languages based on the work carried out for detection of stressed syllables. Further, the applications for which such analysis of stress were carried out were automatic speech recognition (ASR) and speaker recognition. 
Also the speech corpus analyzed for the study was, in most cases, a biased one; for example, emotional speech was recorded (with happiness, anger, sadness) and used for stress analysis. 
Not much literature is found on stress labeling of syllables from neutral (speaking style) speech. 

In our work, the attributes chosen to quantify stress in speech (whether emotional or non-emotional) 
are pitch, amplitude and duration. 
Hereafter, the combination of these three parameters - pitch, amplitude and duration of a phone/syllable will be 
referred to as PAD\footnote{A syllable / phone has certain PAD means that its
pitch is $P$ Hz, its amplitude is $A$ dB and it exists for $D$ seconds.}.

\section{The Approach}
\label{sec:approach} 

Our approach of using the inherent stress information in speech for 
speaker recognition is as follows. 
We first collect a database of some spoken words in which all the phonemes in a 
language occur. 
We make sure that for a given a phoneme, 
we can find several instances of it in the database. 
For example, a phoneme \texttt{/a/} should occur in atleast 
a few words in the database, 
say, \texttt{/jar/}, \texttt{/ball/}, \texttt{/walk/}, \texttt{/mark/}, \texttt{/hard/} etc. 
These words are recorded, say, $m$ times from say, 
$n$ different speakers at different times of the day over a period of several
days. 
The recorded speech samples are then segmented and phoneme 
labeled\footnote{Segmenting and phoneme labeling 
is a process of marking the starting and ending times of phones uttered in the speech sample. 
This process could be automatic or manual.}. 
We use PRAAT \cite{praat} for manually segmenting and phoneme labeling 
the speech samples. 
We then extract the characteristics, namely, pitch, 
amplitude and duration (PADs) of different phones in 
different words and observe the variation of PADs across different instances 
of phones. 

The expectation is that the pitch (amplitude and duration) contours of 
different instances of the same word
of the same speaker appear more or less identical. This gives us an idea of the `speaking style of the speaker' 
which we are interested to capture. One may also be interested in observing the influence of the adjacent phones 
on the characteristics of a phone (especially vowels); for example, how the 
characteristics of the vowel \texttt{/a/}
change in different phonetic contexts. The idea is to come up with a mechanism 
where one can represent a speech utterance with a stress/accent contour that represents the `speaking style' of a person, 
and to use this information in speaker recognition to recognize a speaker. 
 
The mean of the PADs of different instances of each phone uttered by a speaker 
is calculated. 
For each phone uttered by a speaker, 
we have a corresponding mean PAD vector $PAD_m = [P_m, A_m, D_m]$, where $P_m, A_m$ and $D_m$ are the 
average pitch, average amplitude and average duration 
of different instances (recordings at different timings) of a phone. 
The variance, standard deviation of the parameters of each phone and the percentage deviation of each parameter about the mean are also calculated. 
The deviation range of a parameter of a particular phone is determined.
Now the speech feature database of a speaker contains the mean $PAD$ vectors of all the phones in a given language, 
the variance, standard deviation and the percentage deviation (deviation range) of each parameter (PAD) about the mean.

In the training phase, each speaker has a set of phones and their corresponding
mean $PAD$ vectors. 
In the testing phase, we extract the PAD contours of the sentence uttered by the speaker. 
For a verification system, we construct the corresponding PAD contours of the sentence from 
the training database and calculate the distance between the contours of the uttered sentence and 
that which is constructed from the training database. 
If the difference falls within the `deviation range' then the speaker is genuine, else an imposter. 
For an identification system, we construct the corresponding PAD contours of the sentence from the 
training database for all the speakers and calculate the distance between the contours of the uttered sentence 
and those which are constructed from the training database. 
The identified speaker is the one whose constructed PAD contours have minimum distance from the PAD 
contours of the uttered sentence. Note that the minimum distance should also fall within the deviation range. 

\section{Experiments}
\label{sec:experiments} 

For our initial experimentation, we chose $4$ speakers and recorded from 
each of them, $6$ English words.
Speakers were asked to speak the words multiple number of times to check the 
consistency and range of the variance of the 
characteristics of their speech 
samples over the recordings at different timings. 
The words are chosen such that they have a common phoneme \texttt{/a/} in different phonetic contexts. 
The chosen words are \texttt{ball}, \texttt{car}, \texttt{example}, \texttt{hard}, \texttt{mark} and \texttt{wall}. 

The recorded speech samples are manually segmented and labeled using 
PRAAT \cite{praat}. 
After segmentation of the speech samples recorded from different speakers, the values of  
pitch, amplitude and duration (PADs) are obtained for each instance of the phone \texttt{/a/} from a speaker. 
A three-dimensional vector $PAD_{/a/,S_i} = [P, A, D]$ is formed\footnote{$i$ is the index of the speaker}. 
The values of pitch, amplitude and the duration of a phone are obtained as follows. 
PRAAT software has a provision to get the `pitch listing' of a speech file, which gives the values of the 
pitch for every $k$ ms\footnote{$k$ is adjustable}. 
We obtained the values of pitch for every 10 ms and $P$ is the maximum value among those pitch-values 
that fall within the duration of a phone. 
$A$ is obtained from the `intensity listing' of the PRAAT. 
We obtained the values of amplitude for every 10 ms and $A$ is the maximum value among those intensity-values 
that fall within the duration of a phone. 
The duration, $D$ of each phone is obtained from the phone boundaries (on the time scale) 
established by the manual speech segmentation process.
The PAD values and their corresponding mean, variance, standard deviation (SD) and the \%deviation about the mean are obtained for the 
samples of all the speakers. Values of the above mentioned parameters 
for one set of speech samples of a male speaker and are listed in 
Table \ref{tab:pad}.

\begin{table}
\caption{PAD statistics of the phone \texttt{/a/} recorded from a male speaker, Sampling frequency: 8 kHz}
\label{tab:pad}
\centering
\begin{tabular}{|l|l|l|l|} \hline
Word &	Duration (sec)&	Amplitude (dB)&	Pitch (Hz)\\
\hline\hline
\texttt{ball}	& 0.178	& 52.787 &	110.239\\
\texttt{car}	& 0.170 &	47.400 &	113.772\\
\texttt{example}	& 0.128 &	50.657&	116.262\\
\texttt{hard}	&0.205 &	52.001&	127.806\\
\texttt{mark}	 &0.124 &	50.629 &	118.914\\
\texttt{wall}	&0.159 &	50.618 &	110.233\\
\hline			
Mean	&0.161	&50.682&	116.205\\
Variance&	0.0008&	2.825	&36.518\\
SD & 0.0282	&1.681	&6.043\\
\%deviation&	17.554&	3.316 &	5.200 \\
\hline
\end{tabular}
\end{table}

PADs are collected in a similar manner for the sets of speech samples 
recorded from the other speakers.  
As mentioned in Section \ref{sec:approach}, we form the mean $PAD$ vectors $PAD_m = [P_m, A_m, D_m]$ corresponding to the 
different phones of a particular language for each speaker. We compute the other parameters namely, 
variance, standard deviation and percentage deviation about the mean also for the different phones (in a language) 
uttered by each speaker.  
 
After this process, we have the corresponding mean PAD vectors for each phone for each speaker. 
For speaker recognition applications, we extract the PAD contours of the test sentence uttered by the speaker. 
\begin{itemize}
\item For speaker verification system, the corresponding PAD contours of the sentence are constructed by concatenating the 
already available mean PAD values from the training database of the corresponding speaker. 
We can then calculate the Euclidean distance between the PAD contours of the uttered sentence and 
that which is constructed from the training database. If the difference falls within the `deviation range' calculated above, 
then the speaker is declared to be genuine, else an imposter. 
\item For speaker identification, we obtain the PAD contours of the sentence uttered by a speaker. 
We then construct the corresponding PAD contours of the same sentence from the 
training database, for all the speakers. 
We then calculate the Euclidean distance between the contours of the uttered sentence 
and those which are constructed from the training database. 
The identified speaker is the one whose constructed PAD contours have minimum distance from the PAD 
contours of the uttered sentence. Note that, another criteria is that the minimum distance 
should also fall within the deviation range. 
\end{itemize}

\begin{figure}
\centering
\includegraphics[width=0.45\textwidth]{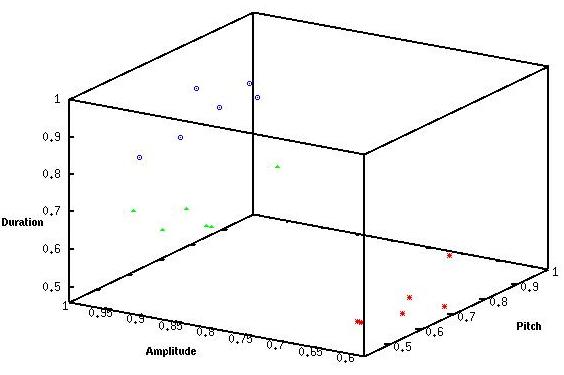}
\caption{3 dimensional PAD vectors of the phone \texttt{/a/} for three different speakers; The values of P, A and D are normalized with respect to the maximum value}
\label{fig:pad}
\end{figure}

An an initial check of the proposed idea, 
we plotted the three-dimensional PAD vectors obtained for different instances of the phone \texttt{/a/}
uttered by three different speakers in the three dimensional space. We found that the PAD vectors of the same phone uttered by 
there different speakers are well separated (see Figure \ref{fig:pad}). This gives us a hope that we can record speech samples 
of all the phones from different speakers construct the training database completely and add the additional feature set to our existing speaker recognition system.

Along with the PADs which are identified as features for stress in speech, we are also collecting 
other features like formants, number of cycles per second for further analysis. 
We hope that this analysis enables us to strongly parametrize the speaking style of a speaker 
and thus use this as an efficient feature for speaker recognition.  

\section{Conclusion}
\label{sec:conclusions}

Speaker recognition systems show a degraded performance when the speaker is 
under 
stress or emotion. 
We made an assumption that there is inherent stress or emotion related 
characteristics
present in spoken speech of a person all the time; in addition the stress
related features are consistent.
Using this assumption as the base, we proposed a mechanism that enables 
the use of 
inherent stress in speech (speaking style) for speaker recognition. We propose
that the stress
related features be used in addition to the regular features used for speaker
recognition. 
We identified $3$ features which capture stress in speech, namely, pitch, amplitude and duration. 
We experimentally observe that the PAD vectors of the similar phones of a speaker are 
close to each other in the three dimensional space 
confirming our assumption that the way a speaker stresses different syllables
in their speech is unique to themselves.
Having observed experimentally that our assumptions are valid, we further proceed to construct a training database of 
PAD values of all the phones in a language for several speakers and incorporate the proposed mechanism in our speaker recognition 
system. 


\bibliographystyle{IEEEtran}
\bibliography{stress_sr}

\end{document}